\documentclass[a4paper]{jpconf}
\usepackage{graphicx}

\newcommand{\cR}{\mathcal{R}}

\begin{document}
\bibliographystyle{iopart-num}
\title{Flavour constraints on the Aligned Two-Higgs-Doublet Model and CP violation}

\author{Paula Tuz\'{o}n}

\address{Departament de F\'{i}sica Te\`{o}rica, IFIC, Universitat de Val\`{e}ncia - CSIC,
  \\
   Apt. Correus 22085, E-46071 Val\`{e}ncia, Spain.}

\ead{paula.tuzon@ific.uv.es}

\begin{abstract}
\noindent
The Aligned Two-Higgs-Doublet Model (A2HDM) \cite{Pich:2009sp} describes a particular way of enlarging the scalar sector of the Standard Model (SM), with a second Higgs doublet which is aligned to first the one in flavour space. This implies the absence of flavour-changing neutral currents (FCNCs) at tree level and the presence of three complex parameters. Within this general approach, we analyze the charged Higgs phenomenology, including $CP$ asymmetries in the $K$ and $B$ systems \cite{Jung:2010ik}.
\end{abstract}

\section{Introduction}
The two-Higgs-doublet model (2HDM) is one of the simplest extensions of the SM, only based on the enlargement of the scalar sector by one more doublet. With this small assumption a rich phenomenology is provided, making the model very interesting not only by itself but also as part of some other extensions of the SM. Without any other model-building constraint, the structure of the Yukawa Lagrangian results in 
\begin{eqnarray} 
-\mathcal L_Y  =     \frac{\sqrt{2}}{v}  \left\{   \bar{Q}_L'  (M_d'  \Phi_1 + Y_d' \Phi_2) d_R'  +  \bar{Q}_L'  (M_u' \tilde{\Phi}_1  + Y_u' \tilde{\Phi}_2)u_R'  +  \mbox{} \bar{L}_L'  (M_l' \Phi_1  + Y_l' \Phi_2)l_R'\; +\; \mathrm{h.c.}  \right\} \, .
\end{eqnarray}
where $ \bar{Q}_L'$ and $\bar{L}_L' $ are the left-handed quark and lepton doublets respectively and $f'_R$ ($f=u,d,l$) the right-handed fermions. The scalar fields are defined in the \emph{Higgs basis}
\begin{eqnarray*}
\Phi_1 = \left[ \begin{array}{c}  G^+ \\ \frac{1}{\sqrt{2}} (v+S_1 + i G^0)    \end{array}    \right] \; , \qquad \Phi_2 = \left[ \begin{array}{c}  H^+ \\ \frac{1}{\sqrt{2}} (S_2 + i S_3)    \end{array}    \right] \; ,
\end{eqnarray*}
where only the first doublet acquires a vacuum expectation value (VEV), $v$, and contains the Goldstone bosons, $G^{\pm}$ and $G^0$. The physical degrees of freedom of the scalars are five, two charged fields, $H^{\pm}$, and three neutrals that need to be rotated to be mass-eigensates, $\{ S_1, S_2, S_3 \} \rightarrow \{ h, H, A \}$. $\tilde{\Phi}_{1,2}(x)\equiv i \tau_2\Phi_{1,2}^*$ are the charge-conjugated scalar fields and $M'_f$ and $Y'_f$ the corresponding Yukawa matrices. Since each right-handed fermion is coupled to two unrelated matrices that in general cannot be diagonalized simultaneously, dangerous FCNC interactions are automatically generated when going to the mass-eigenstate Lagrangian. To avoid (or suppress) these FCNCs, which are strongly constrained by the experiments, many models have been developed from the general 2HDM. A new approach based on the alignment of the Yukawa matrices in flavour space was presented in \cite{Pich:2009sp}. It opens an alternative where FCNCs are absent at tree level and, in addition, the presence of three complex parameters preserves the possibility of having new $CP$ violating sources, which is not possible in other models.

\section{The Aligned two-Higgs-doublet model}\label{athdm}

The alignment condition in flavour space means 
\begin{eqnarray}
Y'_d = \varsigma_d \; M'_d \; , \qquad Y'_u= \varsigma_u^* \; M'_u \; ,  \qquad Y'_l = \varsigma_l \; M'_l \; ,
\end{eqnarray}
with $\varsigma_f$ arbitrary complex numbers. These conditions imply that $Y'_{f}$ are not arbitrary anymore but proportional to $M'_f$ so they can be simultaneously diagonalized: $Y_{d,l} = \varsigma_{d,l} M_{d,l}$ and $Y_u = \varsigma_u^* M_u$. In terms of mass-eigenstate fields, the Yukawa Lagrangian takes then the form
\begin{eqnarray}
-\mathcal L_Y  = \frac{\sqrt{2}}{v} H^+ \left\{ \; \bar{u}  \left[  \varsigma_d V M_d \mathcal P_R - \varsigma_u M_u V \mathcal P_L  \right]   d + \varsigma_l \,\bar{\nu} M_l \mathcal P_R l \; \right\} 
+ \frac{1}{v} \,\sum_{\varphi, f} \varphi^0_i y^{\varphi^0_i}_f \; \bar{f}\;  M_f \mathcal P_R  f 
\; +\;  \mathrm{h.c.} \; ,
\end{eqnarray}
where $V$ is the CKM matrix, $\mathcal P_{R,L}\equiv \frac{1\pm \gamma_5}{2}$ and the neutral couplings $y^{\varphi^0_i}_f$ are given in \cite{Pich:2009sp}. This Lagrangian has the following features: all fermionic couplings are proportional to the mass matrices and the neutral Yukawas are diagonal in flavour. The only flavour-changing structure is the matrix $V$, appearing in the charged current part of the quark sector, like in the SM. There are three new complex parameters, $\varsigma_f$, encoding all the possible freedom allowed by the alignment conditions. They are universal (flavour blind), do not depend on the scalar basis (contrary to the usual $\tan \beta$), recover in some limits all $\mathcal Z_2$-type models, and their phases introduce new sources of $CP$ violation without tree-level FCNCs. This fact represents a counterexample to the very well established idea that the only way of having new $CP$ violation in the electroweak sector of a 2HDM is breaking flavour conservation in neutral current interactions. 

\subsection{Radiative corrections}

The alignment condition is not directly protected by any symmetry, therefore quantum corrections could induce some misalignment generating small FCNCs, that are suppressed by the corresponding loop factors. Nevertheless, the flavour structure of the A2HDM strongly constraints the possible FCNC interactions. The Lagrangian is invariant under flavour dependent phase transformations of the fermion mass eigenstates ($f=u,d,l,\nu$, $X=L,R$, $\alpha^{\nu,L}_i = \alpha^{l,L}_i$), $f_X^i(x)\,\rightarrow \, e^{i \alpha^{f,X}_i}\, f_X^i(x)$ while $V$ and $M_f$ transform like $V_{ij}\,\rightarrow \,e^{i \alpha^{u,L}_i} V_{ij}\, e^{-i\alpha^{d,L}_j}$ and $M_{f,ij}\,   \rightarrow   \,e^{i \alpha^{f,L}_i} M_{f,ij}\, e^{-i\alpha^{f,R}_j}$. Due to this symmetry, lepton flavour violation is zero to all orders in perturbation theory, while in the quark sector the $V$ matrix remains the only source of flavour-changing phenomena. The possible FCNC structures, $\bar u_L F_u^{nm} u_R$ and $\bar d_L F_d^{nm} d_R$, are of the type $F_u^{nm} = V (M_d^{\phantom{\dagger}} M_d^\dagger)^n V^\dagger (M_u^{\phantom{\dagger}} M_u^\dagger)^m M_u^{\phantom{\dagger}}$ and $F_d^{nm}=V^\dagger (M_u^{\phantom{\dagger}} M_u^\dagger)^n V (M_d^{\phantom{\dagger}} M_d^\dagger)^m M_d^{\phantom{\dagger}}$,\ or similar structures with additional factors of $V$, $V^\dagger$ and
quark mass matrices. Therefore, at the quantum level the A2HDM provides an explicit implementation of  the popular Minimal Flavour Violation (MFV) scenarios \cite{Chivukula:1987py, Hall:1990ac, Buras:2000dm, DAmbrosio:2002ex, Cirigliano:2005ck, Kagan:2009bn, Buras:2010mh, Trott:2010iz}, but allowing at the same time for new $CP$ violating phases. 

Using the renormalization group equations \cite{Cvetic:1998uw, Ferreira:2010xe, Braeuninger:2010td}, one finds that the one-loop gauge corrections preserve the alignment, and the only FCNC operator is \cite{Jung:2010ik}
\begin{eqnarray}
\mathcal L^{1Loop}_{\mathrm{FCNC}} &=& \frac{C(\mu)}{4\pi^2 v^3}\; (1+\varsigma_u^*\varsigma_d^{\phantom{*}})\; \sum_i\, \varphi^0_i(x)   \times\nonumber\\
 &\times& \left\{
(\cR_{i2} + i\,\cR_{i3})\, (\varsigma_d^{\phantom{*}}-\varsigma_u^{\phantom{*}})\; 
\left[\bar d_L\, F_d^{01} \, d_R\right] -  (\cR_{i2} - i\,\cR_{i3})\, (\varsigma_d^*-\varsigma_u^*)\;
\left[\bar u_L\, F_u^{10} \, u_R\right] \right\}
 \nonumber \\  &+&\;  \mathrm{h.c.} \; ,
\end{eqnarray}
which of course vanishes in all $\mathcal Z_2$-type models. It is suppressed by $m_qm_{q'}/v^3$ and quark-mixing factors, which implies interesting effects in heavy quark systems like $\bar{s}_L b_R$ and $\bar{c}_L t_R$.

\section{Phenomenology}\label{pheno}

One of the most important features of the A2HDM is the presence of a charged Higgs. In \cite{Jung:2010ik} we analyzed the most relevant flavor-changing processes that are sensitive to charged-scalar exchange and determined the corresponding constraints on the model parameters. We discussed tree-level processes, $\tau \rightarrow \mu/e$, $P^-\rightarrow l^- \nu_l$ and $P\rightarrow P' l^- \bar\nu_l$, where $P$ is a pseudoscalar meson, and loop-induced processes, $\Delta M_{B_s}$, $\epsilon_K$, $Z \rightarrow b \bar b $ and $\bar B \rightarrow X_s \gamma$. 

\begin{figure}
\centering{
\includegraphics[width=5.5cm]{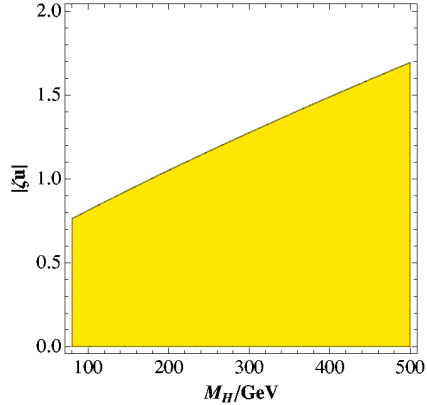}
\caption{\it 95\% CL constraints from $\epsilon_K$. \label{epsk}}}
\end{figure}

\begin{figure}[tb]
\begin{center}
\begin{tabular}{cc}
\includegraphics[width=5.3cm]{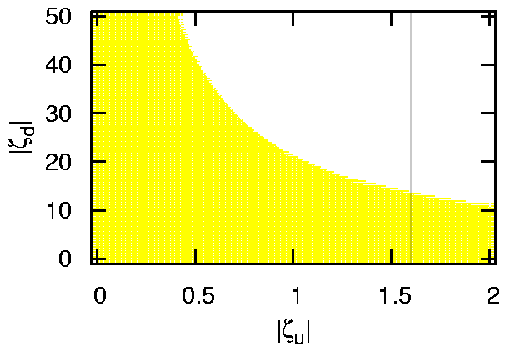} &
\includegraphics[width=5.6cm]{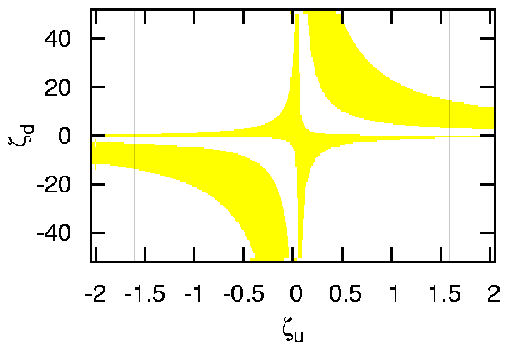} \\
\end{tabular}
\caption{\label{ud} \it
Constraints on $\varsigma_u$ and $\varsigma_d$ from $\bar{B} \rightarrow X_s \gamma$, taking $M_{H^\pm}\in[80,500]$~GeV. The white areas are excluded at $95\%$ CL. The black line corresponds to the upper limit from $\epsilon_K, Z\to\bar{b}b$ on $|\varsigma_u|$.
In the left panel, the relative phase has been varied in the range $\varphi \in [0,2\pi]$. The right panel assumes real couplings.}
\end{center}
\end{figure}

Pure leptonic decays give a direct bound on $|\varsigma_l|/M_{H^{\pm}} \leq 0.40$ GeV$^{-1}$ at $95\%$ CL. Combining this and the information from the other tree-level processes discussed in \cite{Jung:2010ik}, bounds on $\varsigma_u \varsigma_l^*/M_{H^{\pm}}^2$ and $\varsigma_d \varsigma_l^*/M_{H^{\pm}}^2$ parameter space were obtained. Figure \ref{global} shows the resulting limits at $95\%$ CL. These limits are rather weak, allowing the model to fit the experiments in a wide range of its parameter space. Thus, the A2HDM results in a more versatile model than other two-Higgs-doublet models.
\begin{figure}\label{global}
\begin{center}
\includegraphics[width=5.5cm]{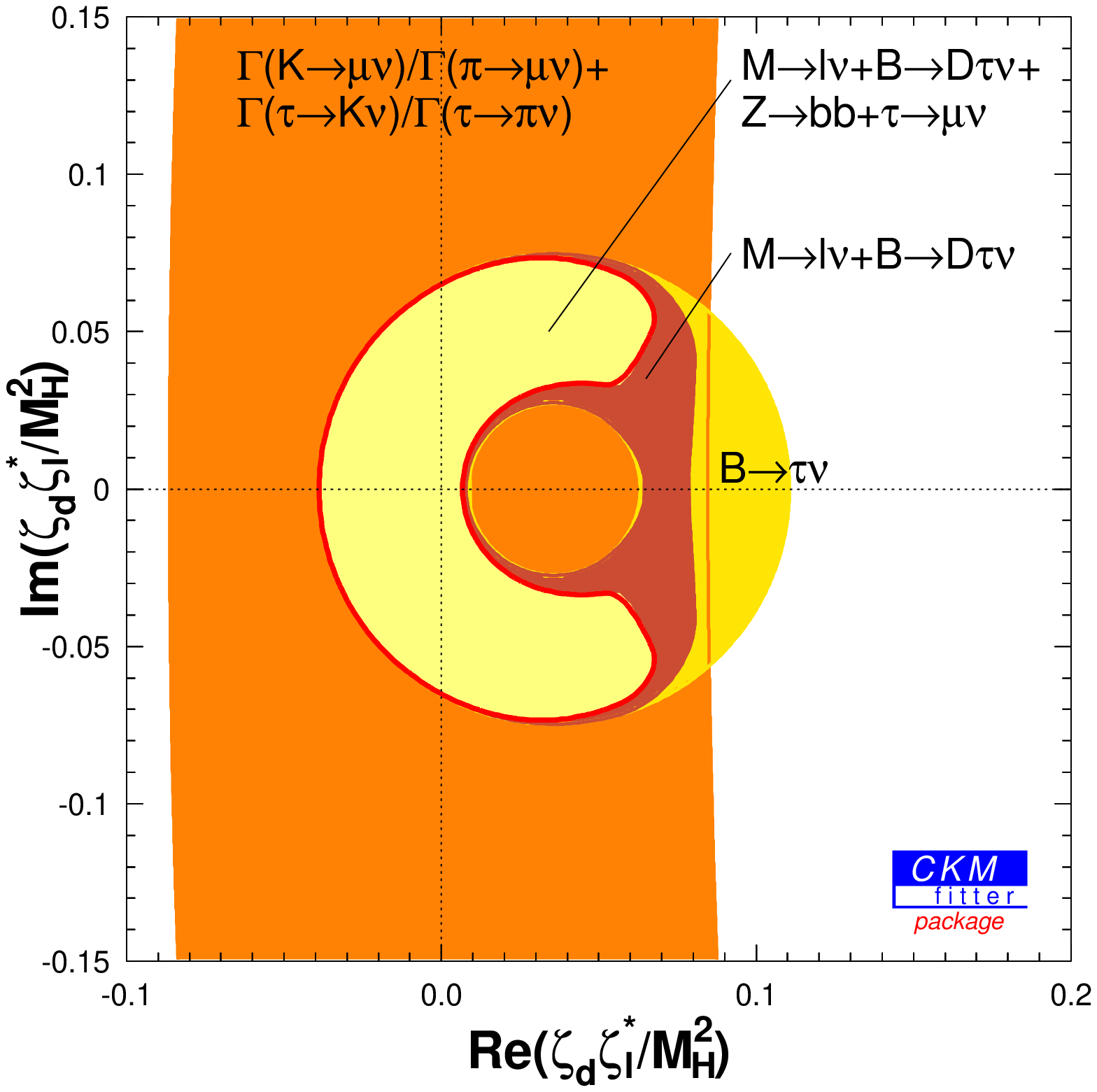} \qquad \includegraphics[width=5.5cm]{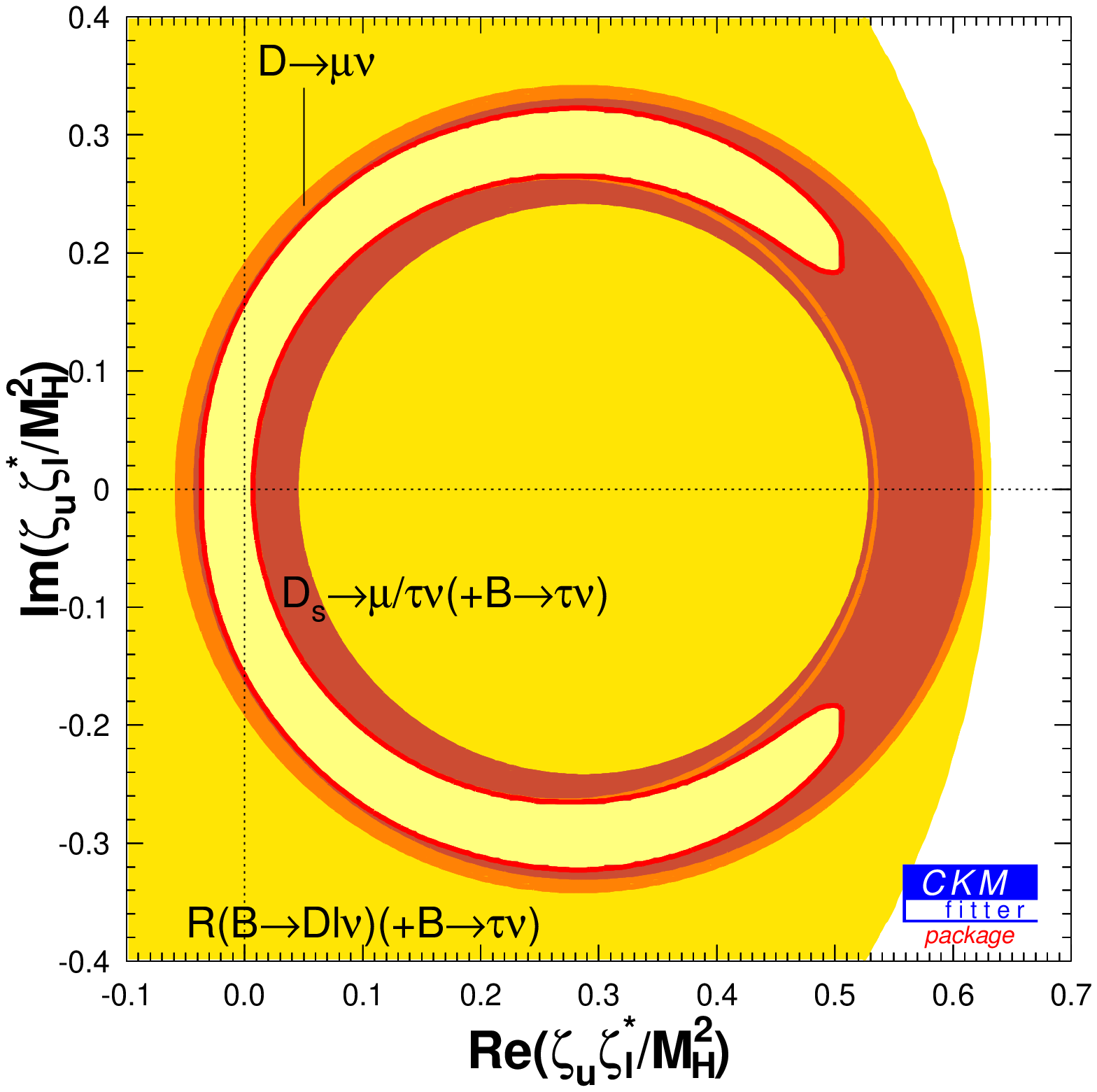}
\caption{\it $\varsigma_d\varsigma_l^*/M_{H^{\pm}}^2$ (left) and $\varsigma_u\varsigma_l^*/M_{H^{\pm}}^2$ (right) in the complex plane, in units of $GeV^{-2}$, constrained by leptonic and semileptonic decays. The inner yellow area shows the allowed region at 95\% CL, in the case of $\varsigma_d\varsigma_l^*/M_{H^{\pm}}^2$ using additional information.\label{global}}
\end{center}
\end{figure}
Loop-induced processes offer direct bounds on $\varsigma_u$ and $\varsigma_d$. No significative constraints on $\varsigma_d$ can be found from $\Delta M_{B_s}$, $\epsilon_K$ and $Z \rightarrow b \bar b$, because $\varsigma_u$-terms are enhanced by the top mass in comparison to $\varsigma_d$-terms. However, quite strong bounds on $\varsigma_u$ are found. Figure \ref{epsk} shows the allowed $|\varsigma_u| - M_{H^{\pm}}$ parameter space at 95$\%$ CL for values of $\varphi \in [0,2\pi]$ and $\varsigma_d \in [0,50]$, given by $\epsilon_K$, which is the most constraining. 
$\bar B \rightarrow X_s \gamma$ gives information in both $\varsigma_u$ and $\varsigma_d$. Figure \ref{ud} shows the resulting constraints on $\varsigma_u - \varsigma_d$ plane for complex (left) and real (right) couplings. $M_{H^{\pm}}$ is scanned over the range $[80,500]$ GeV while $\varphi \in [0,2\pi]$ in the complex case. The main conclusions coming from this figure are the following: First, large couplings of opposite sign are forbidden in the real case. Second, the role of a non-vanishing relative phase is to allow some parameter space which is excluded in the real case. And finally, the product $|\varsigma_d \varsigma_u^*| \leq 20$. Again, we find weaker limits than in other models, showing that the A2HDM gives more possibilities to accomplish the experimental constraints. 

\section{$CP$ violation} \label{cp}

\begin{figure}
\centering{
\includegraphics[width=9cm]{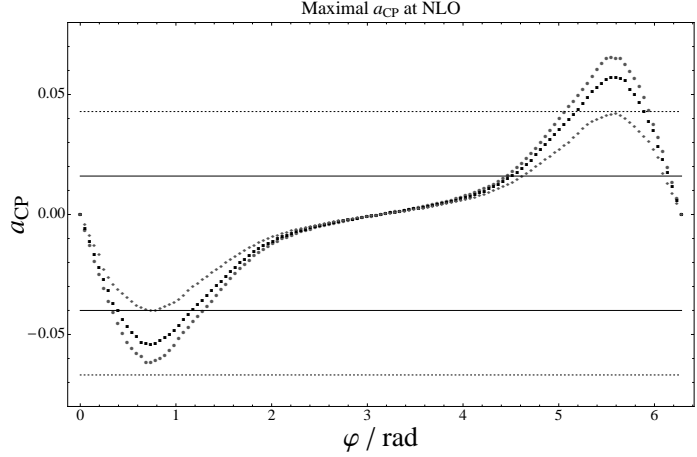}
\caption{\it Maximal $a_{CP}$ over the relative phase $\varphi$ at NLO for $M_{H^{\pm}}\in [80,500]$ GeV, $|\varsigma_u|\in [0,2]$ and $|\varsigma_d|\in[0,50]$, taking into account the experimental constraint on $\bar{B}\rightarrow X_s \gamma$ branching ratio at $95\%$ CL. The three curves correspond to the maximal $a_{CP}$ at $\mu_b=2,2.5,5$ GeV (outer, center and inner respectively), the minimal $a_{CP}$ (black) is always zero independently on the scale. The dotted (continuous) horizontal lines denote the band of the experimental $a_{CP}$ at $1.96\sigma$ ($1\sigma$). \label{acpbsgamma}}}
\end{figure}

The $\bar{B}\rightarrow X_s \gamma$ decay is known to be very sensitive to new physics because the SM prediction for the $CP$ rate asymmetry ($a_{CP}$) is tiny. Requiring that the experimental branching ratio should be correctly reproduced (at $95\%$ CL), we find the results shown in figure \ref{acpbsgamma}. We see that the maximal asymmetry is compatible with the experimental measurement at $95\%$ CL within the scale dependence of the prediction. Then, we conclude that the A2HDM prediction for this observable does not give new constraints on the parameter space that are not already given by the branching ratio, although it reaches the experimental bound. Therefore, a precise measurement of this asymmetry and a more accurate calculation in both SM and 2HDM parts of the branching ratio would be very interesting. 

\begin{figure}
\centering{
\includegraphics[width=8cm]{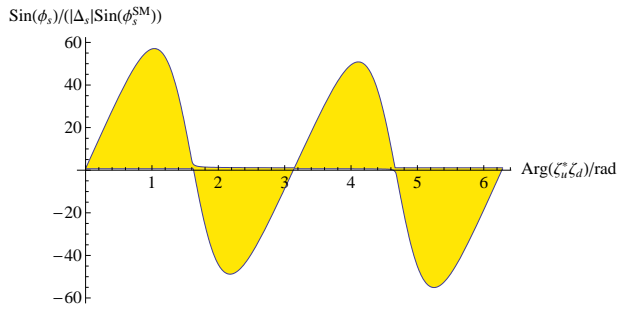}
\caption{\it Dependence of $a_{CP}^{A2HDM}/a_{CP}^{SM}$ on $\varphi$, constraining $|\varsigma_u^* \varsigma_d| \leq 20$, for $M_{H^{\pm}}$, $|\varsigma_u|$ and $\varsigma_d$ scanned in the same ranges as in figure \ref{acpbsgamma}. \label{acpbmix}}}
\end{figure}

Some months ago, the D0 experiment measured an enhanced like-sign dimuon charge asymmetry \cite{Abazov:2010hv} in the $B_s$ system incompatible with a purely SM rate. In \cite{Jung:2010ik} we concluded that although the D0 central value is quite unlikely, it is possible to accommodate an enhanced $a_{CP}$ within the A2HDM. Figure \ref{acpbmix} shows how large is this enhancement ($a_{CP}^{A2HDM}/a_{CP}^{SM}$) depending on the relative phase $\varphi$, where the bound on the product $|\varsigma_u^* \varsigma_d| < 20$ coming from $\bar{B}\rightarrow X_s \gamma$ has been taken into account. From the plot we see that the asymmetry can be enhanced even 60 times compared to the SM. The preferred negative sign of $a_{sl}^s$ constrains $\varphi \in [\pi/2,\pi], [3\pi/2, 2\pi]$. A large asymmetry requires large $|\varsigma_d|$ values and small values for the charged Higgs mass. 

Due to the different structure of the $b\rightarrow s \gamma$ and $B_s^0-\bar{B}_s^0$ amplitudes, the enhanced asymmetry in each case corresponds to different regions in the A2HDM parameter space, giving complementary information on the relative phase.

\section{Conclusions} \label{conclusions}

The A2HDM provides a powerful realization of a 2HDM with no tree-level FCNCs and three complex parameters $\varsigma_f$. These parameters are new sources of $CP$ violation, flavour blind, scalar basis independent and recover in some limits all the models implemented by discrete $\mathcal Z_2$ symmetries. This parametrization allows for more freedom and accomplishes all the experimental constraints. Some misalignment can come from quantum corrections, only as MFV structures which are under control and just occur in the quark sector. Processes involving a charged Higgs give information on the parameters $\varsigma_f$, although they result in weaker limits compared to the usual scenarios with $\mathcal Z_2$, where the freedom introduced by the $\varsigma_f$ phases does not exist. The $CP$ asymmetries generated in $\bar{B}\rightarrow X_s \gamma$ and in the $B_s$ systems within the A2HDM enhance the SM prediction in complementary regions. The predicted asymmetry for $\bar{B}\rightarrow X_s \gamma$ does not give new bounds on the parameter space compared to the branching ratio, regarding this process, a more precise measurement and a complete calculation reducing the theoretical error are essential. On the other hand, if the experiments confirm a large asymmetry in $B_s$, it could be perfectly accommodated in this framework.

\section*{Acknowledgements}
This work has been done in collaboration with Martin Jung and Antonio Pich. It has been supported in part by the EU MRTN network FLAVIAnet [Contract No. MRTN-CT-2006-035482] and by MICINN, Spain
[FPU No. AP2006-04522, Grants FPA2007-60323 and Consolider-Ingenio 2010 Program CSD2007-00042 --CPAN--].

\section*{References}
\bibliographystyle{iopart-num}
\bibliography{bibliography}

\end{document}